\documentclass[twocolumn]{article}

\pdfoutput=1

\usepackage{amsmath,graphicx,url}

\usepackage[a4paper,left=1.8cm,right=1.8cm,top=2cm,bottom=2.5cm]{geometry}

\usepackage{ifpdf}
\ifpdf
    \usepackage[pdftex,breaklinks=true]{hyperref}
\else
    \usepackage[breaklinks=true]{hyperref}
\fi

\newlength\imgwidth
\begin{document}

\title{An Improved DC Recovery Method from AC Coefficients of
DCT-Transformed Images\thanks{Companion Web page: \protect\url{http://www.hooklee.com/default.asp?t=AC2DC}.}}

\author{Shujun Li\textsuperscript{1}, Junaid Jameel Ahmad\textsuperscript{1}, Dietmar Saupe\textsuperscript{1} and C.-C. Jay
Kuo\textsuperscript{2}}

\date{%
\textsuperscript{1} Dept. of Computer \& Information Science\\
University of Konstanz, Germany\\
\textsuperscript{2} Ming Hsieh Dept. of EE\\
University of Southern California, USA}

\maketitle

\begin{abstract}
Motivated by the work of Uehara \textit{et al.} \cite{USO:AC2DC:IEEETIP2006}, an improved method to recover DC coefficients from AC coefficients of DCT-transformed images is investigated in this work, which finds applications in cryptanalysis of selective multimedia encryption. The proposed under/over-flow rate minimization (FRM) method employs an optimization process to get a statistically more accurate estimation of unknown DC coefficients, thus achieving a better recovery performance. It was shown by experimental results based on 200 test images that the proposed DC recovery method significantly improves the quality of most recovered images in terms of the PSNR values and several state-of-the-art objective image quality assessment (IQA) metrics such as SSIM and MS-SSIM.
\end{abstract}

\section{Introduction}

The discrete cosine transform (DCT) is an orthogonal transform with sub-optimal performance in terms of de-correlation efficiency \cite{ANR:DCT:IEEETComp1974}. Since DCT is easier to implement than the optimal Karhunen-Lo\`{e}ve transform (KLT), it has been widely used in signal and image processing applications, especially for lossy image and video compression. Many well known image and video coding standards, including JPEG and MPEG-1/2/4/H.26x, are based on 2D block DCT \cite{Shi-Sun:IVC2nd:Book2008}.

Among all DCT coefficients, the first one, called the DC (direct current) coefficient, plays the most important role. It represents the average intensity of a block and carries most of the energy and the perceptual information. Actually, DC coefficients of all blocks form a thumbnail version of the original image at a lower spatial resolution. When the block-DCT is applied to image/video compression, DC coefficients often consume more bits than other DCT coefficients called AC (alternative current) coefficients. To give an example, to encode the 8-bit gray-scale Lenna image of size $512\times 512$ using the $8\times 8$ block-DCT and the default quantization table of JPEG \cite[Table~K.1]{JPEG}, we estimate that about 16\% of all coding bits will be used to encode quantized DC coefficients, where the estimation is obtained by calculating entropies of quantized DCT coefficients at different locations. If we apply differential coding to DC coefficients, the ratio will decrease to 14\%, which is still significant.

Due to the importance of DC coefficients, it was argued that encrypting DC coefficients \textit{only} may effectively conceal significant visual information to achieve the goal of light-weight selective encryption \cite{Weng-Preneel:DC-Encryption:SIGMAP2007}. For example, Li \textit{et al.} \cite{Li:PVEA:IEEETCASVT2007} proposed to encrypt DC coefficients to conceal a rough view of MPEG-encoded video sequences to achieve perceptual encryption. Other hybrid image/video encryption schemes using the combination of DC encryption and secret AC permutations were proposed in \cite{Tang:MPEGEncryption:ACMMM96}. However, since secret permutations are not secure against plaintext attacks \cite{Li:SPIC2008}, the hybrid encryption scheme can be downgraded to the DC encryption only.

Since DCT is an orthogonal transform, DC encryption is believed to be secure because DC coefficients are independent of AC coefficients. It was however shown by Uehara, Safavi-Naini and Ogunbona \cite{USO:AC2DC:IEEETIP2006} that most DC coefficients of a DCT-transformed natural image can be approximately recovered from AC coefficients.  Their DC recovery method, called the USO method, exploits two properties of most digital images. First, there is strong correlation between neighboring pixels. Second, the DC coefficient of each block is constrained to an interval defined by the DC-free edition of this block, {\em i.e.,} pixel values calculated only from AC coefficients.

In this work, we propose an improved DC recovery method, which outperforms the USO method in recovery quality. Our idea is to minimize the under/over-flow rate of pixel values during the recovery process. We provide experimental results to demonstrate the superior performance of the improved DC recovery method in terms of PSNR and nine objective image quality assessment (IQA) metrics in the MeTriX MuX Visual Quality Assessment Package \cite{MeTriXMuX}, which includes SSIM \cite{Wang:SSIM:IEEETIP2004}, MS-SSIM \cite{Wang:MS-SSIM:Asilomar2003}, IFC \cite{Sheikh:IFC:IEEETIP2005}, VIF \cite{Sheikh:VIF:IEEETIP2006}, etc.

The rest of the paper is organized as follows.  The USO method is reviewed in Sec.~\ref{sec:review}. The proposed DC recovery method is described in detail in Sec.~\ref{sec:FRM-Method}. Finally, the last section gives concluding remarks and future work.

\section{The USO Method}\label{sec:review}

In this section, we present the implementation detail of the USO method by following \cite{USO:AC2DC:IEEETIP2006}, where some missing details in the original paper are filled in with our best efforts. Since in \cite{USO:AC2DC:IEEETIP2006} it was shown that the USO method is robust against quantization, in this paper we do not consider quantization. To facilitate our discussion, we denote the DC coefficient and the average intensity of an image block $B$ by $\text{DC}(B)$ and $\overline{B}$, respectively. We use the orthogonal form of 2D DCT so that $\text{DC}(B)=N\cdot \overline{B}$, where $N$ is the block size. Given a block $B$, define $B^{(d)}$ as the block derived from $B$ by setting $\text{DC}(B)$ to $d$, \textit{i.e.}, $B^{(d)}=(B-\overline{B})+d/N=B+(d-\text{DC}(B))/N$. The valid range of pixel values is denoted by $[t_{\min},t_{\max}]$.

The USO method is based on the following two properties of most digital images.\\
\textbf{Property~1}\quad\textit{The difference between two neighboring pixels is a Laplacian variate with zero mean and a small variance.} \\
\textbf{Property~2}\quad\textit{The range of pixel values calculated from $B^{(0)}$ (i.e., only from AC coefficients) constrains the value of $\text{DC}(B)$.}

By exploiting Property~1, the unknown DC coefficient of a block can be estimated from its neighboring blocks with known DC coefficients by minimizing the differences of adjacent pixels along the block boundary. Three patterns of adjacent pixels were considered in the USO method: one horizontal or vertical pattern, and two diagonal patterns as shown in \cite[Fig.~2]{USO:AC2DC:IEEETIP2006}. The smoothest pattern (\textit{i.e.}, the one with the smallest gradient) is chosen for DC estimation.

Property~2 can be used to determine a valid range of $\text{DC}(B)$. Since $t_{\min} \leq B=B^{(0)}+\text{DC}(B)/N \leq t_{\max}$, we obtain
\begin{equation}\label{equation:AC2DCRange}
N(t_{\min}-\min(B^{(0)})) \leq \text{DC}(B) \leq N(t_{\max}-\max(B^{(0)})),
\end{equation}
which gives an estimation of $\text{DC}(B)$ with an accuracy defined by $N(t_{\max}-t_{\min}+\min(B^{(0)})-\max(B^{(0)}))$.

Uehara \textit{et al.} took the above two properties into account and proposed the following method to recover DC coefficients of a DCT-transformed image with known AC coefficients. Without loss of generality, we assume the input image is a DC-free image, \textit{i.e.}, an image composed of blocks whose DC coefficients are all zeros.\footnote{In principle, the DC coefficients can be any fixed value. In \cite{USO:AC2DC:IEEETIP2006}, the midpoint of the valid range (1023 for 8-bit gray-scale images) is used. Here, we use zero to simplify our description.}

\textbf{Step~1}: Choose a corner block of the DC-free image as the initial reference block $B_0$ and estimate DC coefficients of other blocks (relative to $\text{DC}(B_0)=0$) by scanning the whole image from $B_0$ to the diagonally opposite corner block. Property~1 is used to estimate the DC of each block from its neighboring block. When there are two neighboring blocks, the two estimates are averaged. Ideally, the output of this step is an image identical with the original image except that the average intensity is darker by $\text{DC}(B_0)/N$.

\textbf{Step~2}: The goal of this step is to adjust the whole image so that the recovered image has the same average intensity as the original one. First, calculate the valid DC range of each block $B_i$ from Eq.~\eqref{equation:AC2DCRange}. Then, for each DC range $[d_{i,\min},d_{i,\max}]$, get a range of the intensity adjustment as follows:
$[d_{i,\min}^*,d_{i,\max}^*]=[(d_{i,\min}-\text{DC}^*(B_i))/N,(d_{i,\max}-\text{DC}^*(B_i))/N]$,
where $\text{DC}^*(B_i)$ denotes the relative DC coefficient of $B_i$ estimated in Step~1. Next, calculate $d_{\min}^*=\max\limits_i(d_{i,\min}^*)$ and $d_{\max}^*=\min_i(d_{i,\max}^*)$. Finally, the whole image is brightened by $(d_{\min}^*+d_{\max}^*)/2$.

\textbf{Step~3}: Repeat Steps~1 and 2 by choosing the four corner blocks as initial reference blocks, and average the images obtained from four different scans to get the final result.

\textbf{Step~4}: Since the estimated DC values in previous steps may not be accurate, the output pixel values in Step~3 may not be in the valid range. Then, a post-processing operation, which includes the re-scaling of the whole image or the adjustment of under/over-flow pixel values only, can be performed.

No specific postprocessing operation in Step~4 was mentioned in \cite{USO:AC2DC:IEEETIP2006}. We found from our experiments that the following post-processing scheme works well.
\begin{itemize}
\item
If the dynamic range of pixel values is larger than $t_{\max}-t_{\min}$, scale the whole image to $[t_{\min},t_{\max}]$;

\item
If the dynamic range of pixel values is not larger than $t_{\max}-t_{\min}$, adjust the average intensity of the image towards $[t_{\min},t_{\max}]$ until all pixel values are within $[t_{\min},t_{\max}]$.
\end{itemize}
The above scheme is therefore adopted in our implementation of the USO method. For the four test images with reported PSNR values in \cite{USO:AC2DC:IEEETIP2006}, we obtained a higher average PSNR value (21.7 dB vs. 20.9 dB) under the above setting to generate data given in \cite[Table~III-2]{USO:AC2DC:IEEETIP2006}.

\section{Proposed DC Recovery Method}\label{sec:FRM-Method}

One major drawback of the USO method is the low accuracy of the DC estimation process for some images. Small estimation errors in Step~1 may propagate to result in a large number of under/over-flows of pixel values. These under/over-flows can also affect Step~2 due to the inaccurately estimated DC ranges of some blocks.

\begin{figure}[!htb]
\centering
\begin{minipage}{\imgwidth}
\centering
\includegraphics[width=\textwidth]{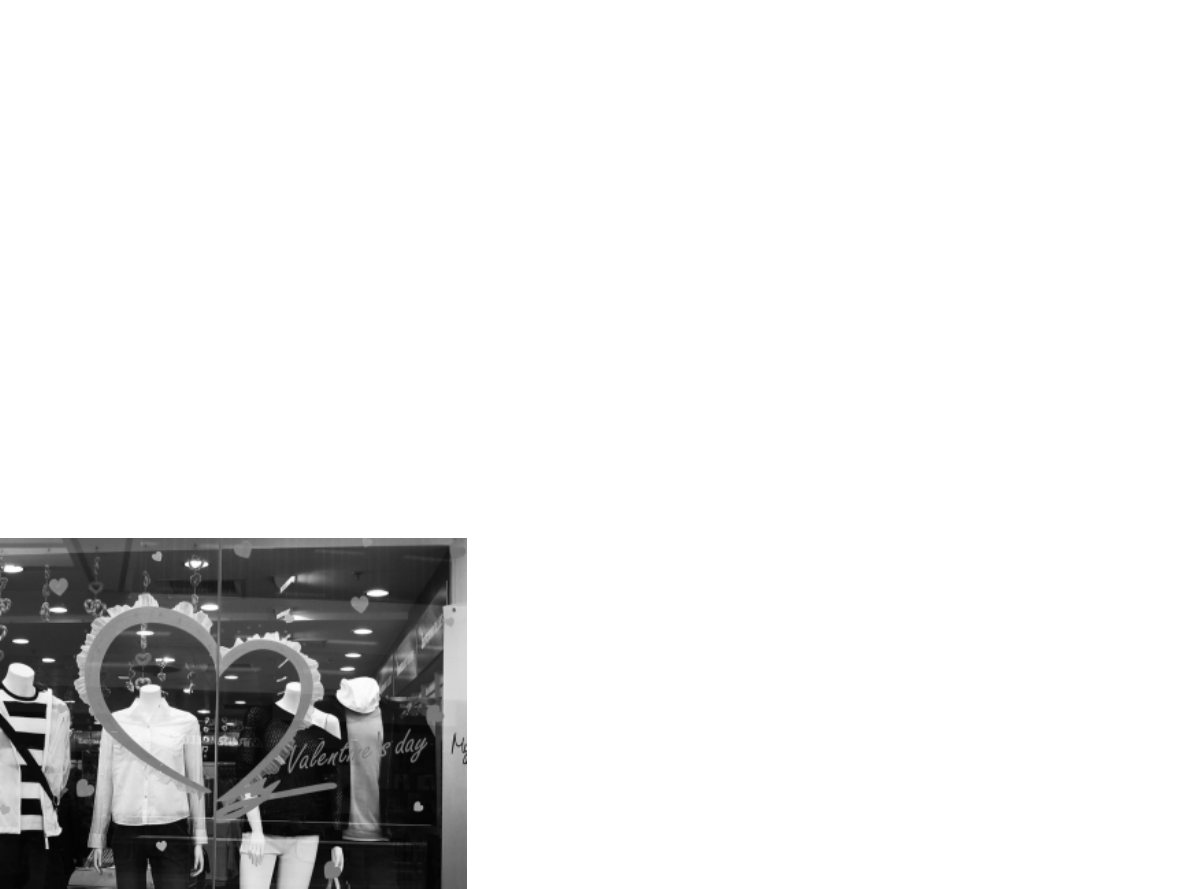} (a)
\end{minipage}
\hfil
\begin{minipage}{\imgwidth} \centering
\includegraphics[width=\textwidth]{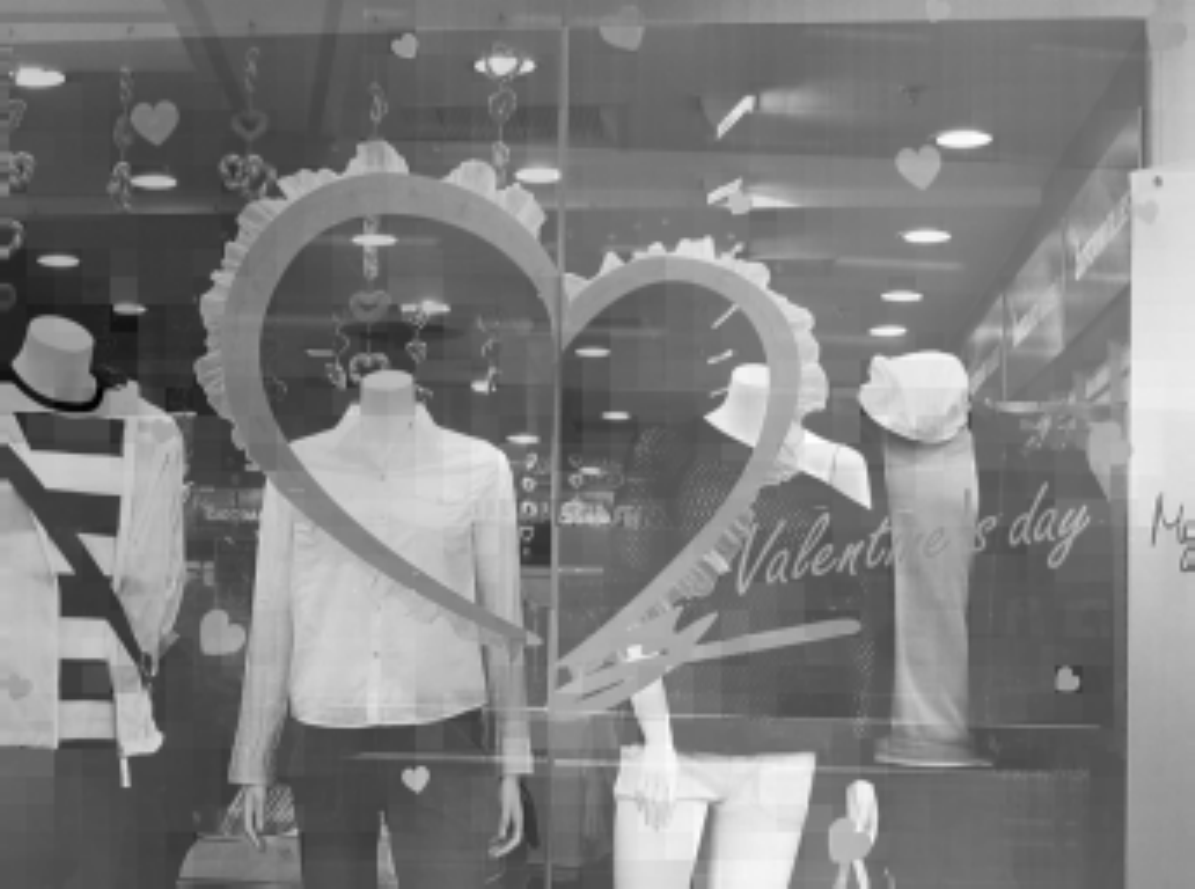} (b)
\end{minipage}
\caption{(a) A test image of size $384\times 256$ and (b) the recovered image using the USO method.}\label{fig:HK_shop}
\end{figure}

\begin{figure}[!htb]
\centering
\begin{minipage}{\imgwidth}
\centering
\includegraphics[width=\textwidth]{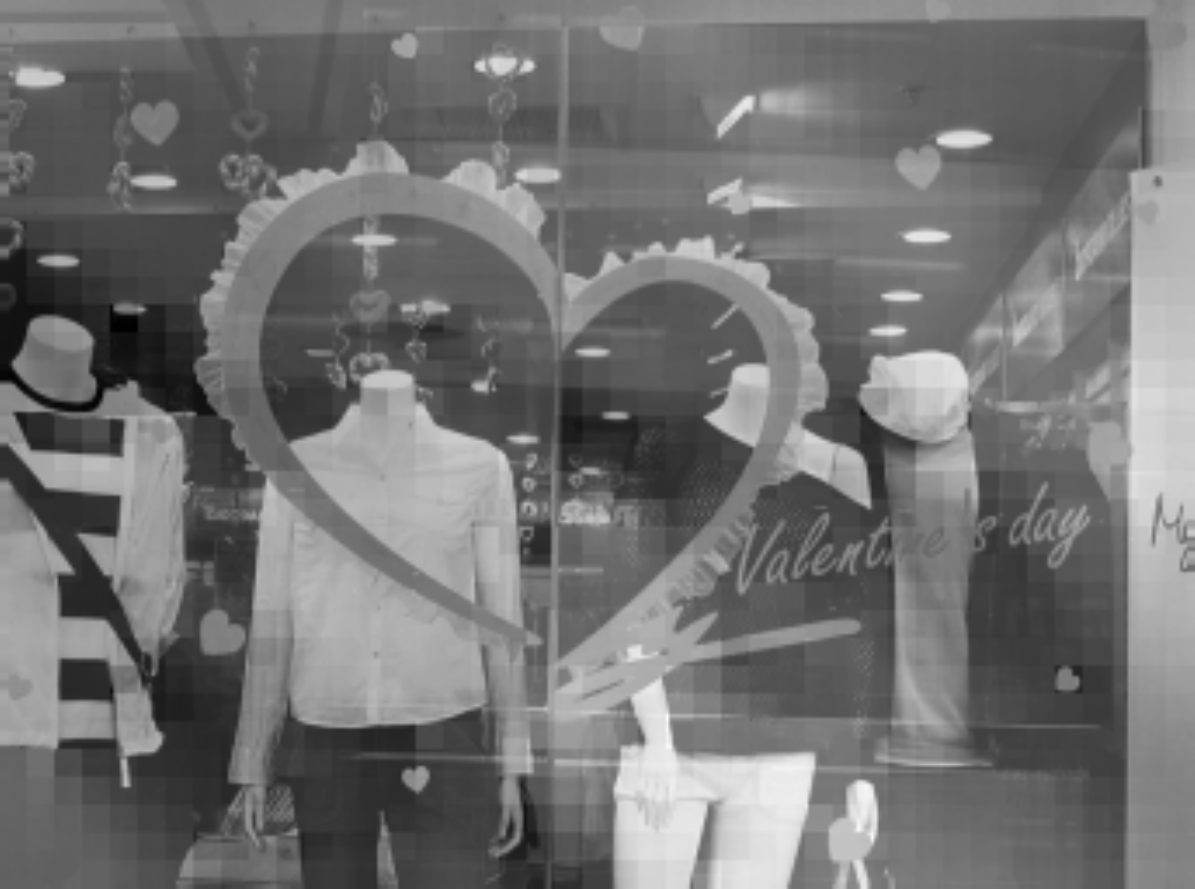} (a)
\end{minipage}
\hfil
\begin{minipage}{\imgwidth}
\centering
\includegraphics[width=\textwidth]{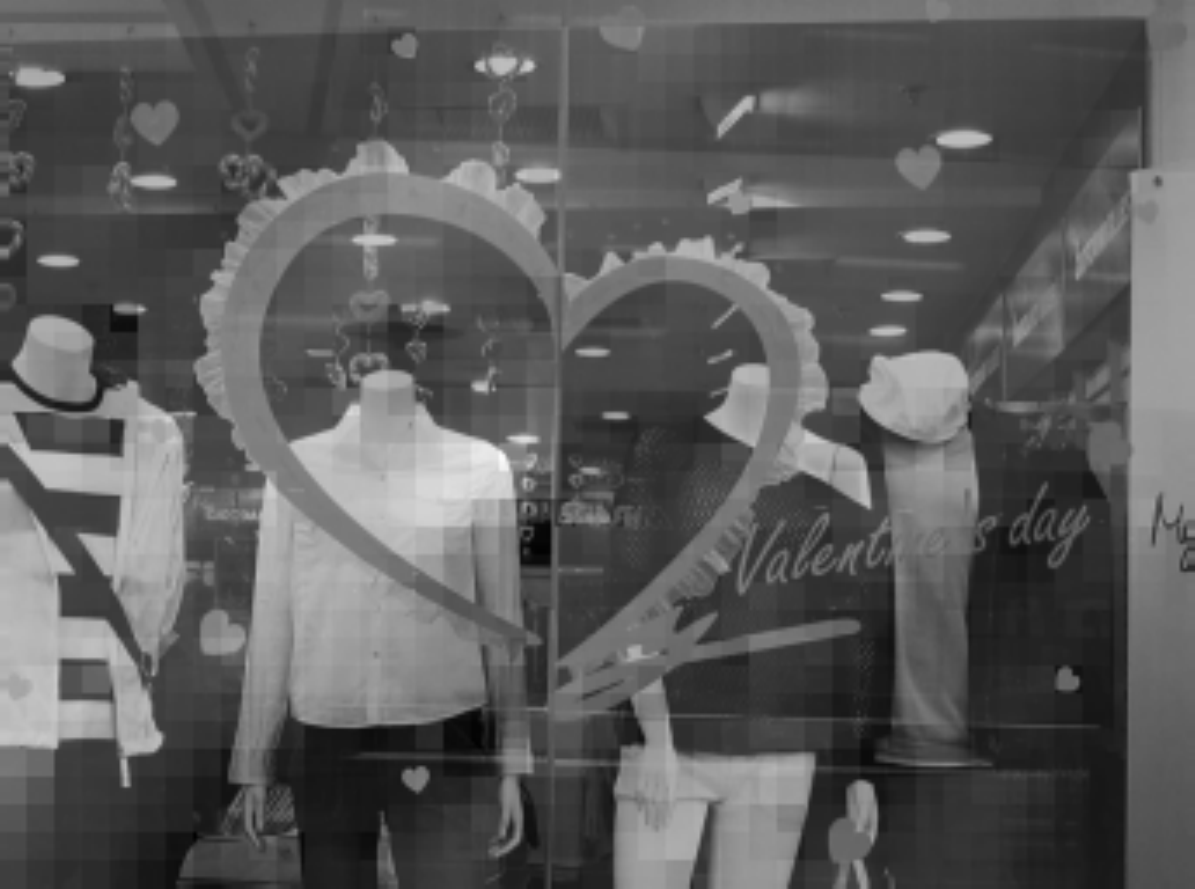} (b)
\end{minipage}\\
\begin{minipage}{\imgwidth}
\centering
\includegraphics[width=\textwidth]{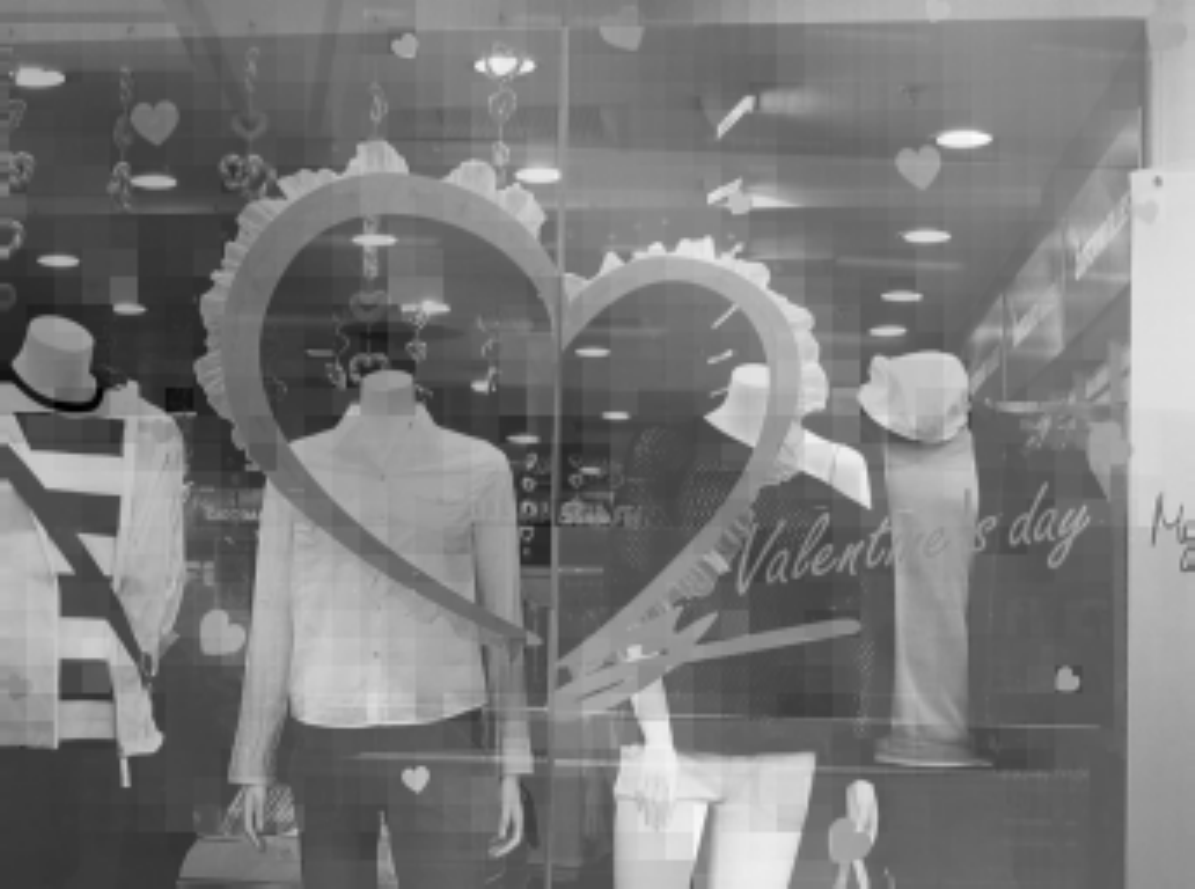} (c)
\end{minipage}
\hfil
\begin{minipage}{\imgwidth}
\centering
\includegraphics[width=\textwidth]{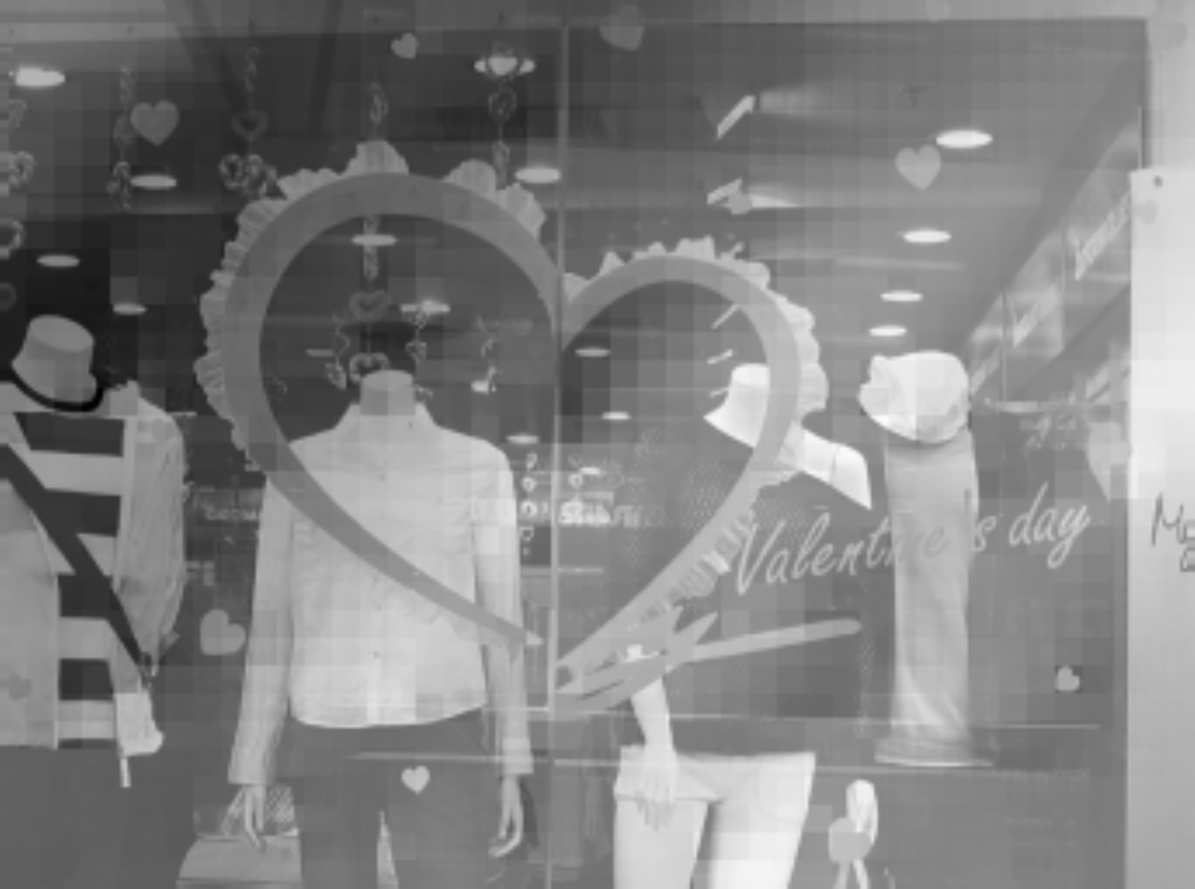} (d)
\end{minipage}
\caption{Four intermediate images recovered from the DC-free edition of the test image Fig.~\ref{fig:HK_shop}(a) using the USO method.}
\label{fig:HK_shop_USO_4Scans}
\end{figure}

To illustrate this problem, we examine the 8-bit gray scale image in Fig.~\ref{fig:HK_shop}(a). The four images obtained from four different scans (after Step~2) are shown in Fig.~\ref{fig:HK_shop_USO_4Scans}(a)--(d). Their pixel value ranges are $[-83.0, 338.0]$, $[-90.3, 345.3]$, $[-92.0, 347.0]$, $[-136.3, 391.3]$, respectively. We see that the error propagation effect is very serious. By averaging the four images, the pixel value range is $[-88.6, 303.0]$, which is still far from the valid one. After scaling the range to $[0, 255]$, we get a better image as shown in Fig.~\ref{fig:HK_shop}(b). However, the quality remains poor with a PSNR value of 14.3, an SSIM score of 0.732 and an MS-SSIM score of 0.711.

\subsection{Under/Over-Flow Rate Minimization (FRM) Method}

To limit the error propagation effect, we apply Property 2 during the relative DC estimation process in Step~1 as follows. For each block, after the relative DC estimation is obtained, we immediately check if the estimated DC coefficient is outside of the valid range. If this occurs, we re-adjust the estimated DC toward the valid range until all pixel values fall in interval $[t_{\min},t_{\max}]$. After that, the scanning process proceeds to the next block. Since the DC-bounding process is done during the scanning process for all blocks, the output of Step~1 will contain \textit{no} underflow or overflow pixel values. Then, Steps~2 and 4 become unnecessary and can be removed.

The modified DC estimation method can effectively limit error propagation. However, it encounters a new problem. That is, it may not be able to recover the DC coefficient of each block accurately if the estimated value of $\text{DC}(B_0)$ is far from the ground truth. This is because the estimates obtained in Step~1 are actually DC coefficients of all blocks \textit{relative} to $\text{DC}(B_0)$. Thus, it is important to find an accurate estimate of $\text{DC}(B_0)$ to ensure that DC coefficients of most blocks are estimated with an acceptable accuracy.

While there is no prior knowledge on the ground truth of $\text{DC}(B_0)$, we can exploit a statistical approach to estimate its value for natural images. It is based on an interesting phenomenon of the run-time DC-bounding process in Step~1: when the estimate of $\text{DC}(B_0)$ is closer to the ground truth, under/over-flows in pixel values tend to occur less frequently. Although some counter examples exist, this observation holds for most tested natural images.\footnote{It is difficult to theoretically analyze the accuracy and sensitivity of the estimation process of $\text{DC}(B_0)$. We conducted experiments on 200 test images to confirm its correctness and efficiency (see Sec.~\ref{sec:Experiments} for more detail).}

\begin{figure}[!htb]
\centering
\begin{minipage}{\imgwidth}
\centering
\includegraphics[width=\textwidth]{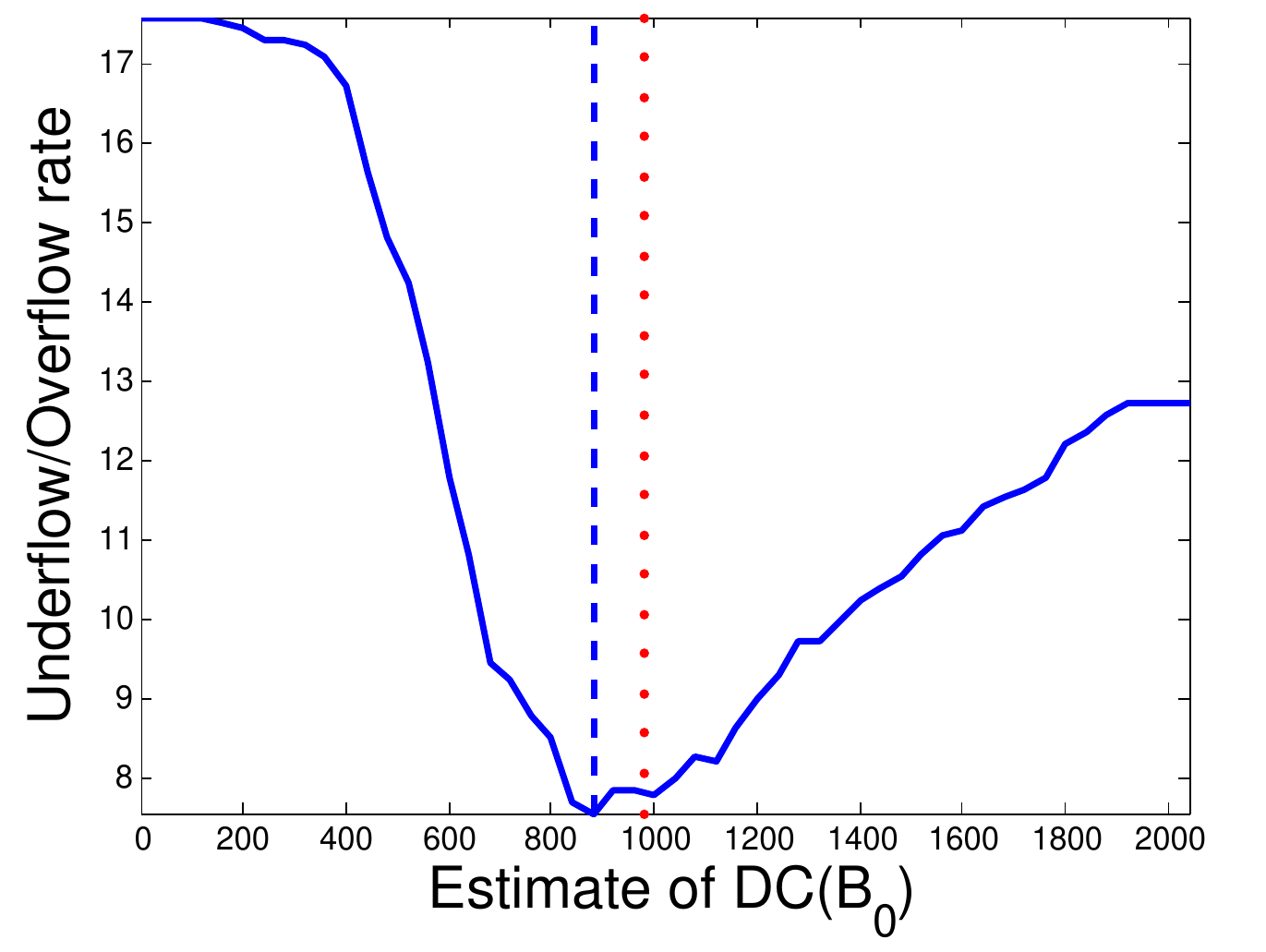}
(a)
\end{minipage}
\hfil
\begin{minipage}{\imgwidth}
\centering
\includegraphics[width=\textwidth]{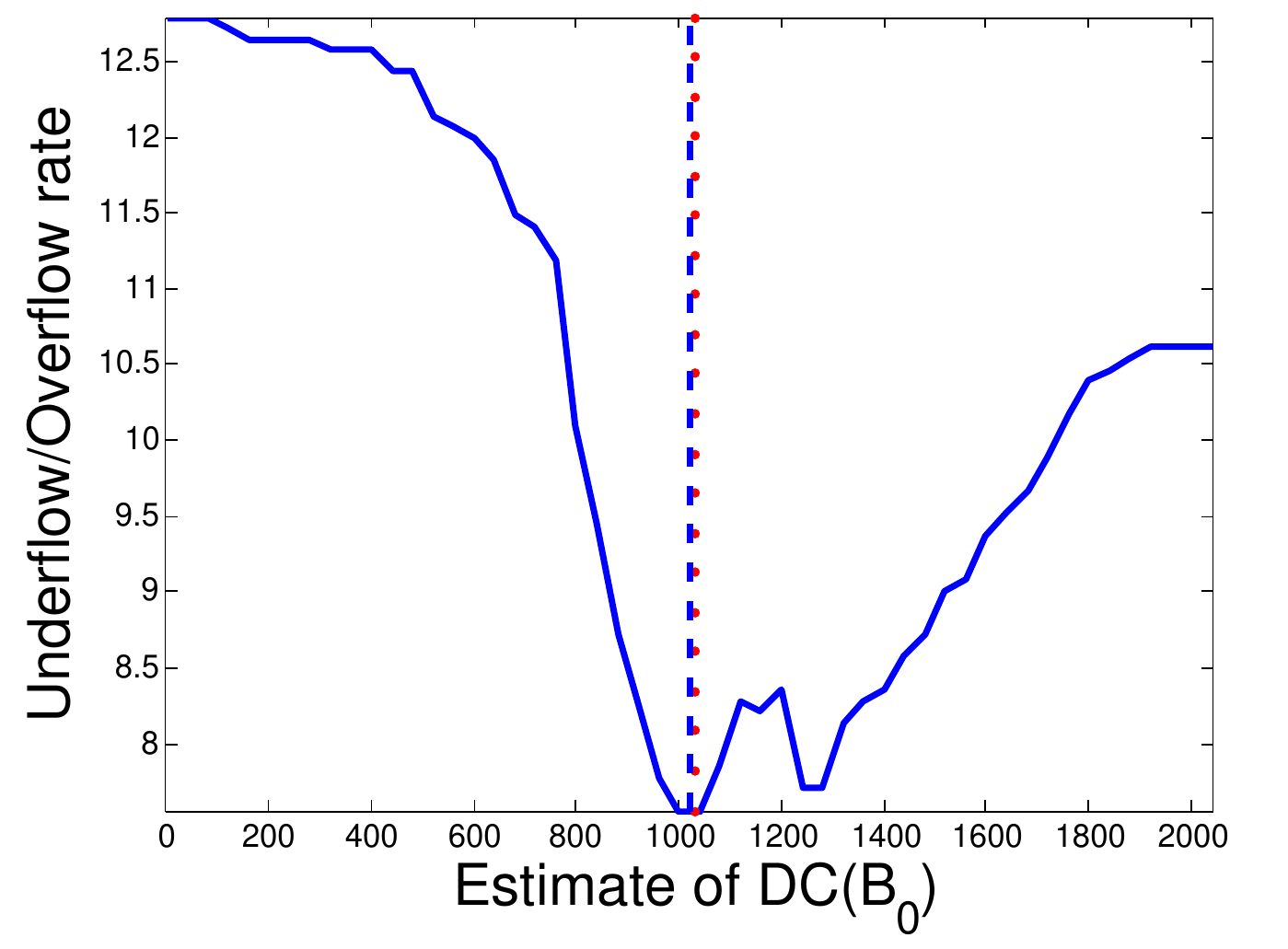}
(b)
\end{minipage}\\
\begin{minipage}{\imgwidth}
\centering
\includegraphics[width=\textwidth]{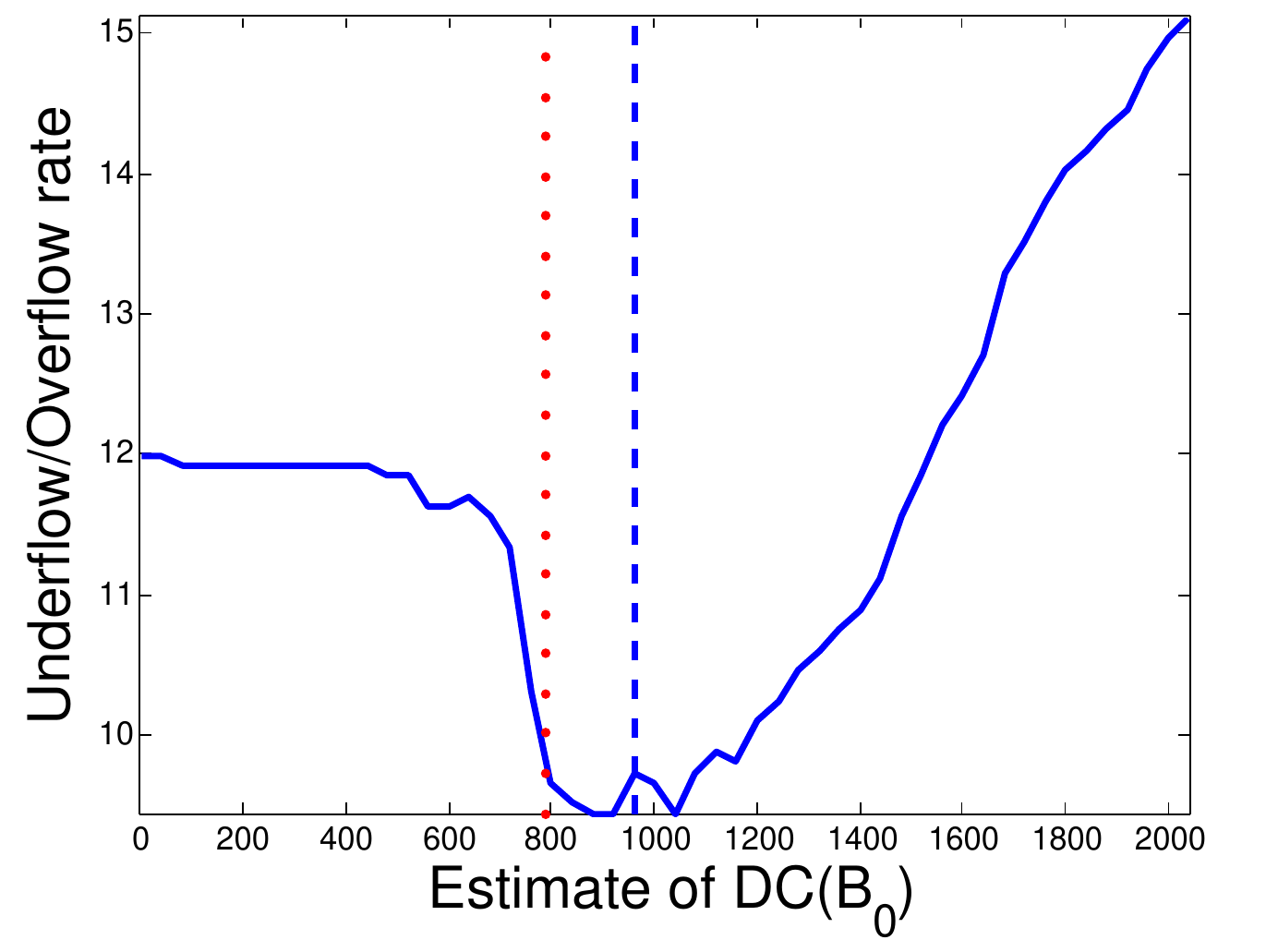}
(c)
\end{minipage}
\hfil
\begin{minipage}{\imgwidth}
\centering
\includegraphics[width=\textwidth]{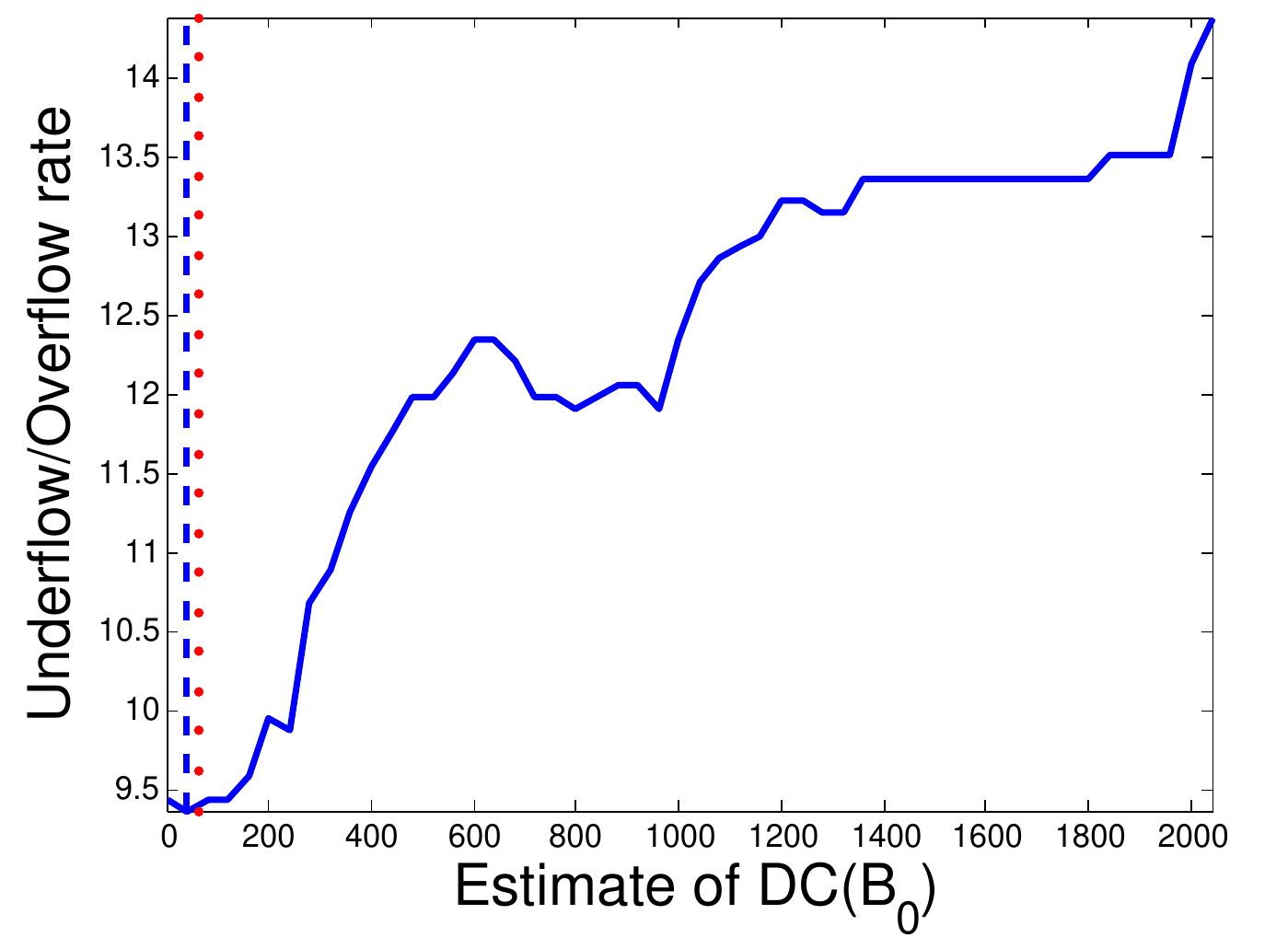}
(d)
\end{minipage}
\caption{The relationship between the under/over-flow rate and the estimate of $\text{DC}(B_0)$ for the test image Fig.~\ref{fig:HK_shop}(a) where Subfigures (a)-(d) give results for Scans~1-4, respectively. } \label{fig:HK_shop_FR_DC0}
\end{figure}

We use an example to illustrate the above observation. For the test image shown in Fig.~\ref{fig:HK_shop}(a), the relationship between the under/over-flow rate and the estimated $\text{DC}(B_0)$ is shown in Fig.~\ref{fig:HK_shop_FR_DC0} for four different scans. The under/over-flow rate is defined as $M_{\text{FR}}/M_B$, where $M_{\text{FR}}$ is the number of blocks with under/over-flow pixel values and $M_B$ is the total number of blocks in the image. In Fig.~\ref{fig:HK_shop_FR_DC0}, the dotted line gives the ground truth of $\text{DC}(B_0)$ while the dashed line gives the $\text{DC}(B_0)$ value that has the minimal under/over-flow rate. When there are multiple points that have the minimal under/over-flow rate, the dashed line shows the mid-point of the range defined by the leftmost and the rightmost points. It is clear that the minimal under/over-flow rate always occurs around the ground truth of $\text{DC}(B_0)$.

\begin{figure}[!htb]
\centering
\begin{minipage}{\imgwidth}
\centering
\includegraphics[width=\textwidth]{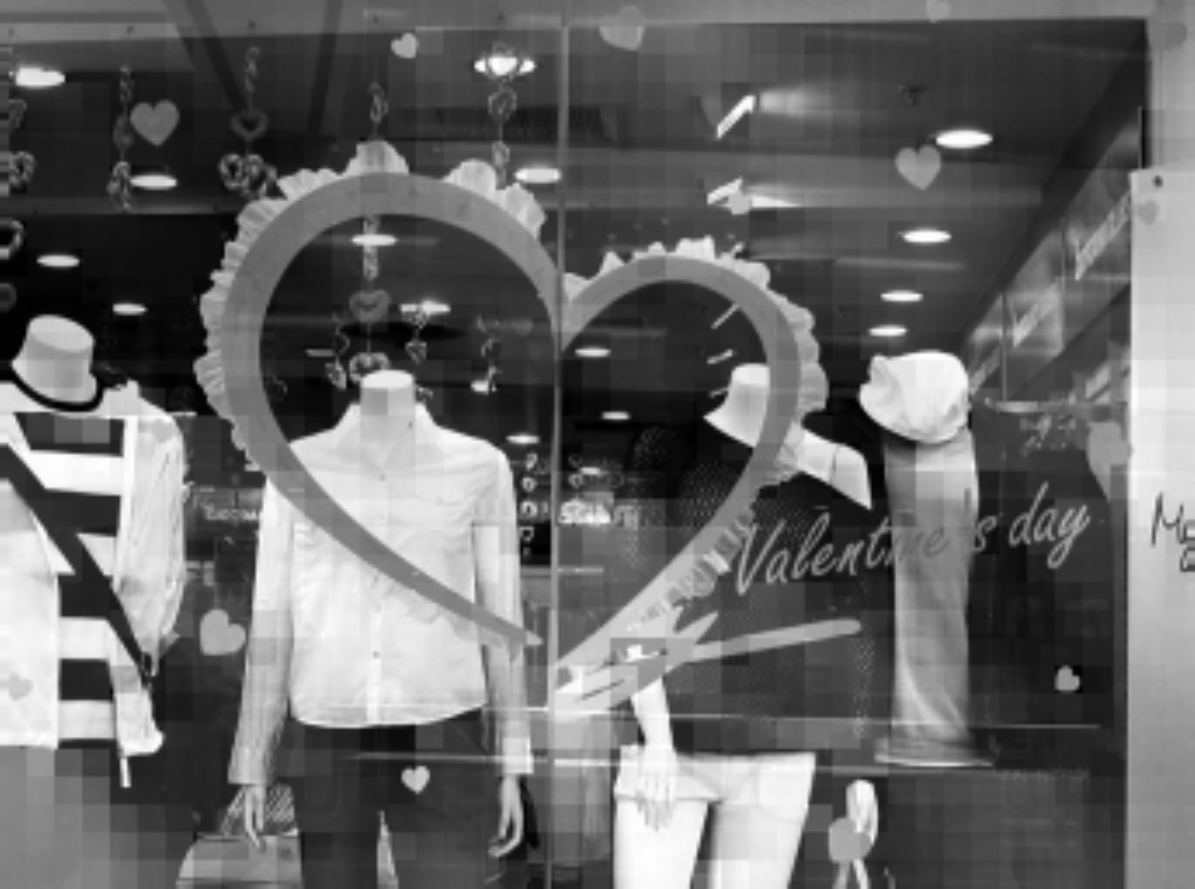}
(a)
\end{minipage}
\hfil
\begin{minipage}{\imgwidth}
\centering
\includegraphics[width=\textwidth]{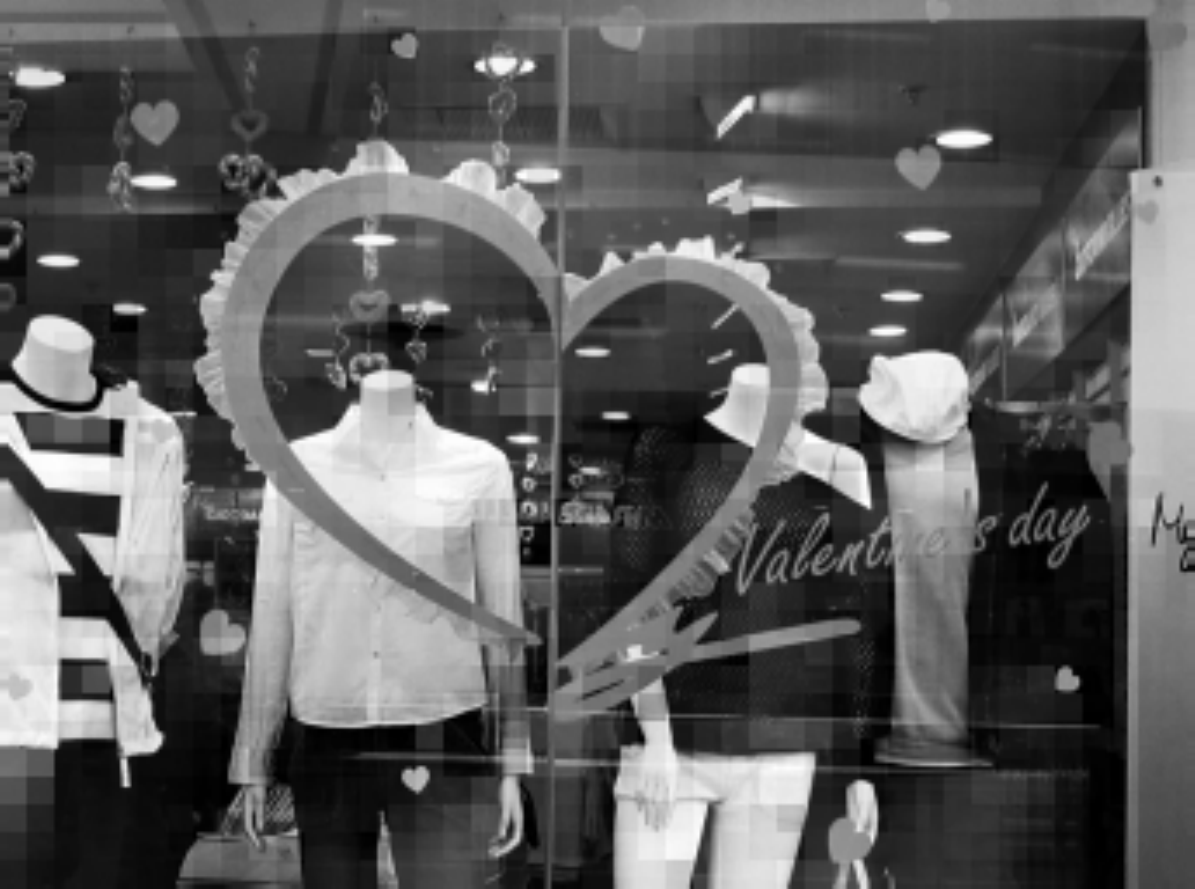}
(b)
\end{minipage}
\begin{minipage}{\imgwidth}
\centering
\includegraphics[width=\textwidth]{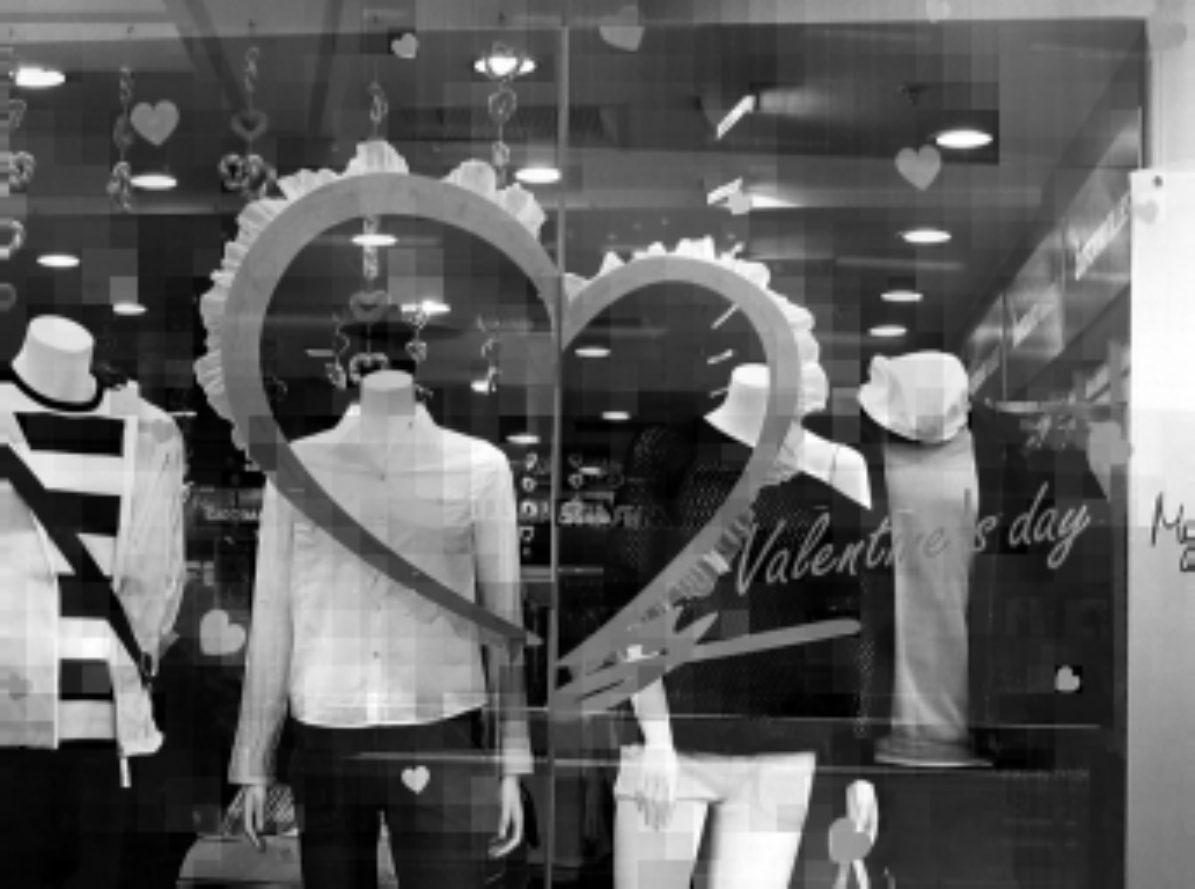}
(c)
\end{minipage}
\hfil
\begin{minipage}{\imgwidth}
\centering
\includegraphics[width=\textwidth]{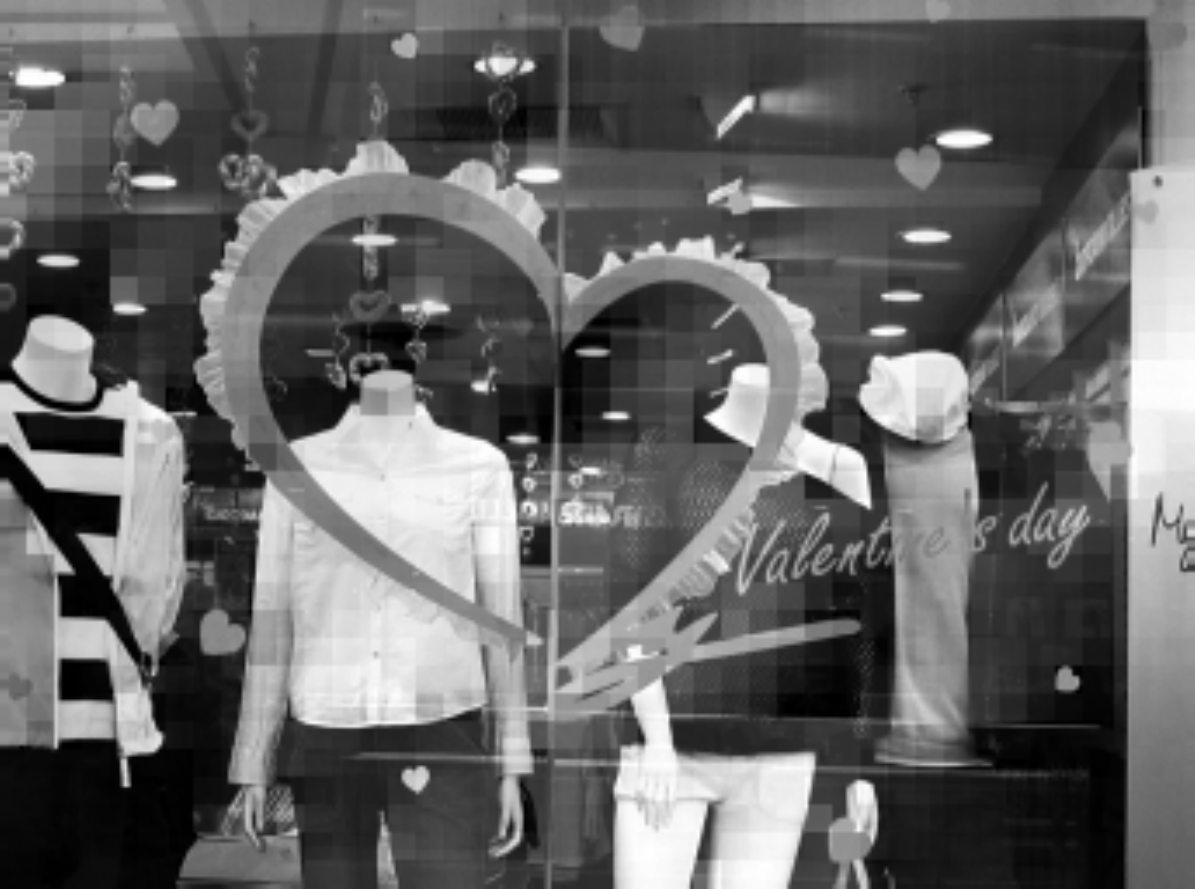}
(d)
\end{minipage}
\caption{The images obtained by minimizing the under/over-flow rate for four scans of the test image in Fig.~\ref{fig:HK_shop}(a), where Subfigures (a)-(d) give results for Scans~1-4, respectively.}
\label{fig:HK_shop_FRM_4Scans}
\end{figure}

\begin{figure}[!htb]
\centering
\includegraphics[width=\imgwidth]{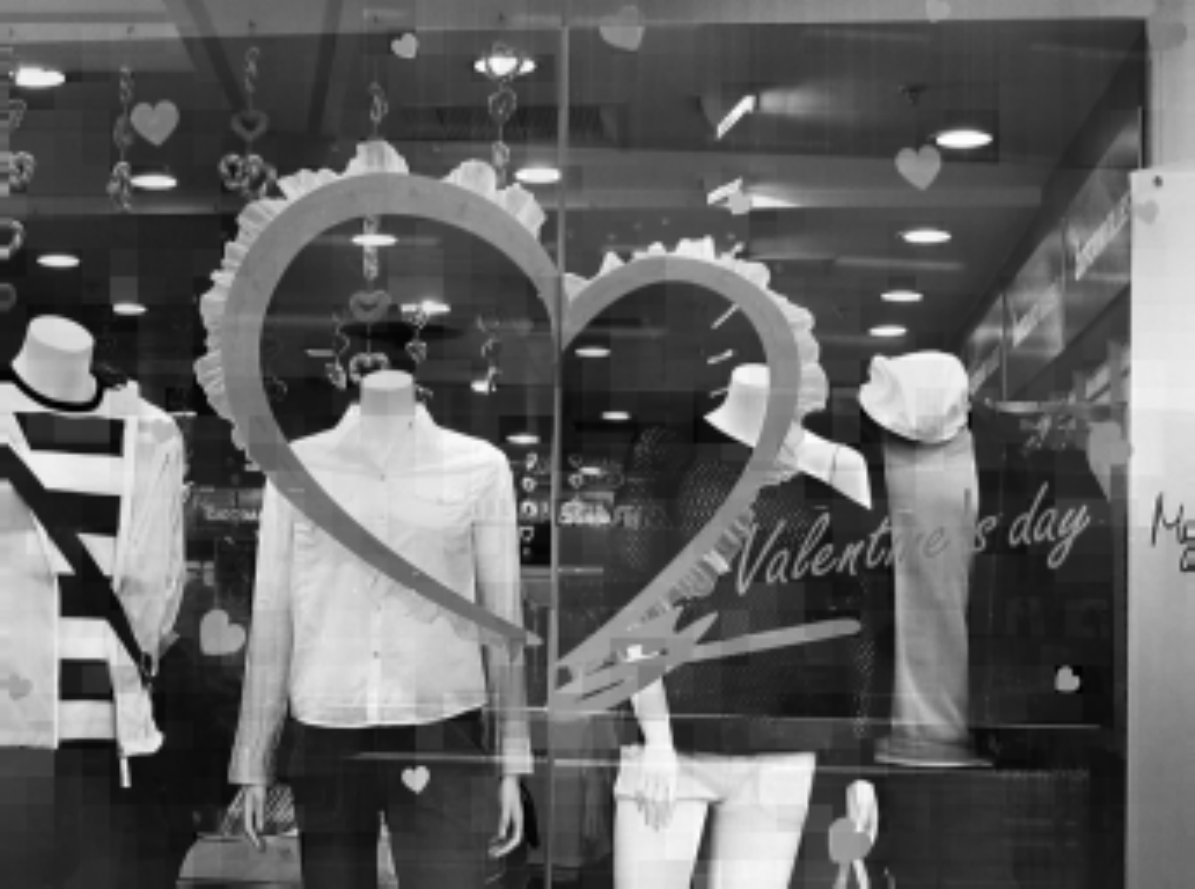}
\caption{The final recovered image using the FRM method from the DC-free edition of the test image in Fig.~\ref{fig:HK_shop}(a).}\label{fig:HK_shop_FRM}
\end{figure}

If the best estimate of $\text{DC}(B_0)$ is used for each scan, we obtain the results in Fig.~\ref{fig:HK_shop_FRM_4Scans}. By comparing Figs.~\ref{fig:HK_shop_FRM_4Scans} with \ref{fig:HK_shop_USO_4Scans}, we see that the proposed FRM method performs significantly better. By averaging the four images obtained by four scans, we show the final recovered image in Fig.~\ref{fig:HK_shop_FRM}. This recovered image has a PSNR value of 23.2, an SSIM score of 0.900 and an MS-SSIM score of 0.924.

\begin{figure*}[!htb]
\centering
\includegraphics[width=\textwidth]{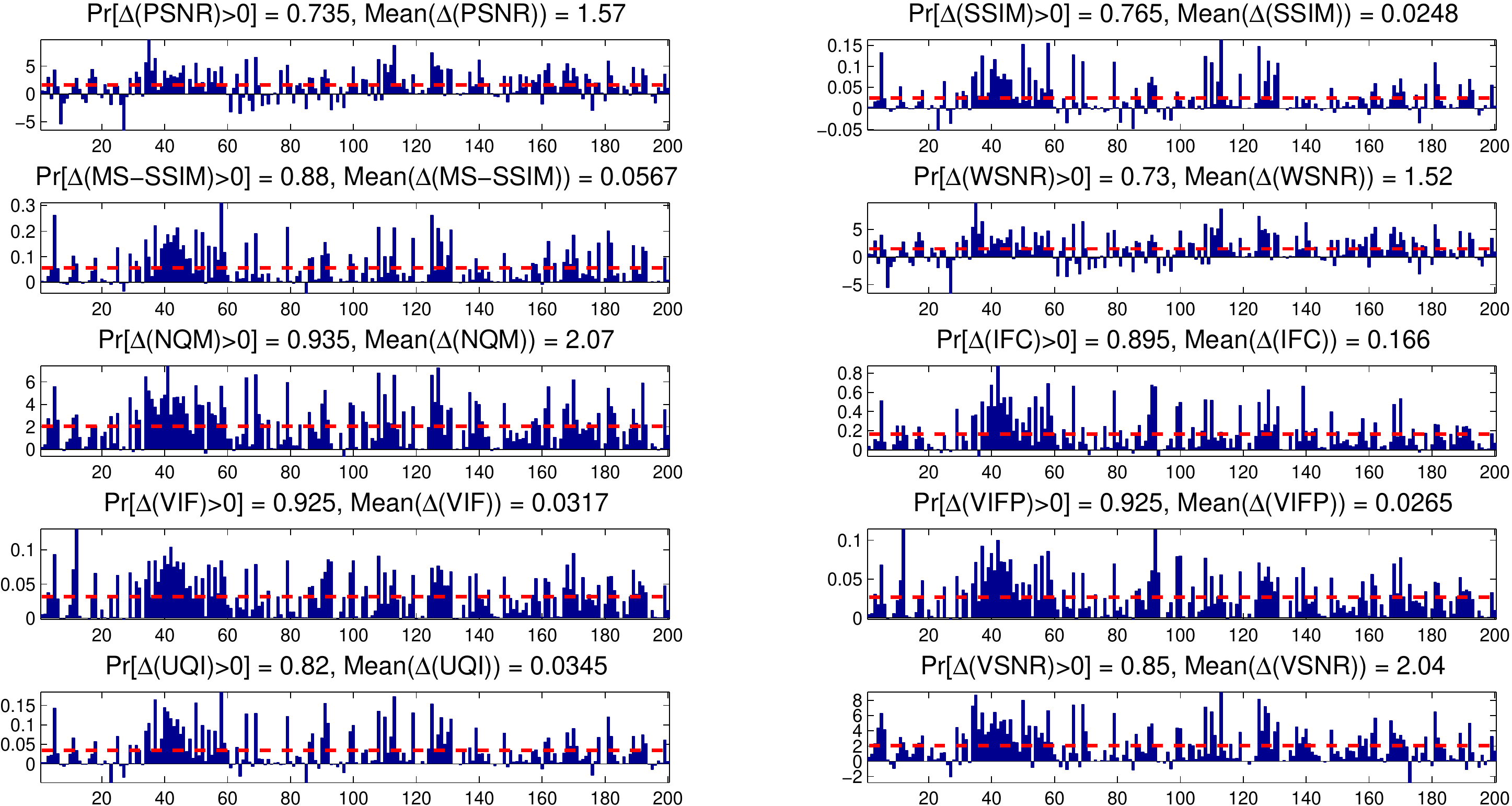}
\caption{The performance improvement of the proposed FRM method over the USO method using different IQA metrics. Note that the range of some metrics (SSIM, MS-SSIM, VIF, VIFP, UIQ) is [0,1], so the mean values corresponding to these metrics are relatively small.}\label{fig:FRM_vs_USO_IQAs}
\end{figure*}

\subsection{Computational Complexity}

The above FRM method is essentially an optimization process of the observable under/over-flow rate during the DC estimation process. To get the best result, an exhaustive search of all possible values of $\text{DC}(B_0)$ could be tried. When the step size is $\Delta$, $N(t_{\max}-t_{\min}+\min(B_0^*)-\max(B_0^*))/\Delta$ values of $\text{DC}(B_0)$ are checked for each block in each scan, and the computational complexity of the FRM method is $O\left(M_BN(t_{\max}-t_{\min}+\min(B_0^*)-\max(B_0^*))/\Delta\right)\leq O\left(M_BN(t_{\max}-t_{\min})/\Delta\right)$. In contrast, the USO method has a complexity of $O(M_B)$, so it is faster than the FRM method.

To reduce the computational complexity, we notice that the relationship between the under/over-flow rate and $\text{DC}(B_0)$ is close to a uni-modal function. Thus, a binary-search strategy can be used to locate the optimal $\text{DC}(B_0)$. At the beginning, the two ending points and the midpoint of the whole valid DC range are searched. Then, the DC range is reduced to be the sub-interval corresponding to the two smaller underflow/overflow rates. This process is repeated until a pre-defined precision is reached. The computational complexity of such a binary search process is less than $O\left(M_B\log_2(N(t_{\max}-t_{\min})/\Delta)\right)$.

Note that the computational complexity of the FRM method is $O(M_B)$ if $\max(B_0^*)-\min(B_0^*)\approx t_{\max}-t_{\min}$, which may happen for some images. In this case, it is as efficient as the USO method.

\subsection{Experimental Results}\label{sec:Experiments}

To verify the effectiveness of the proposed FRM method, we conducted experiments on an image database of 200 test images. Our MATLAB code and the test images are available at \url{http://www.hooklee.com/default.asp?t=AC2DC}. To measure the quality of recovered images, we compare the performance of the proposed FRM method with that of the USO method using the PSNR values and other nine IQA metrics in the MeTriX MuX Visual Quality Assessment Package \cite{MeTriXMuX}. The results are shown in Fig.~\ref{fig:FRM_vs_USO_IQAs}, where the $x$-axis is the image index, and the $y$-axis is the difference of IQA scores and the dashed line shows the mean of the IQA differences of all images. A positive $y$-value means the proposed FRM method outperforms the USO method. We can see that the performance improvement is consistent and significant for most images. Although PSNR and WSNR values become significantly worse for a few images (\textit{e.g.}, Images~7 and 27), their visual quality decreases only slightly or remains nearly the same if measured with other (more accurate) metrics like MS-SSIM and VIF.

\section{Conclusion and Future Work}\label{sec:conclusion}

A method to recover DC coefficients from AC coefficients of DCT-transformed images, called the USO method, was proposed in \cite{USO:AC2DC:IEEETIP2006}. In this work, we proposed an improved DC recovery method and called it the FRM method. Experiments showed the performance improvement of the FRM method over the USO method for 200 test images in terms of the PSNR value and nine different IQA metrics.

There are several ways to further improve the proposed FRM method. For instance, if we have some statistics of the image to be recovered, we may be able to get a better estimate of $\text{DC}(B_0)$ from the minimized underflow/overflow rate. We may define the underflow/overflow rate in a different way, {\em e.g.,} modulating the simple definition with the degree of the underflow/overflow. In addition, more advanced DC prediction and averaging schemes may be used by exploiting the prior knowledge on the underlying image. Another possible improvement is to change the locations of the initial reference blocks and/or the scanning pattern. Theoretically speaking, any block can be the initial reference block and any scanning pattern can be used. It may also be beneficial to have multiple initial reference blocks and scanning patterns. One of the implications of the DC recovery method is that a fewer amount of information about DC coefficients should be encoded in image compression. We will investigate this possibility as well in our future work. Yet another work of interest is to compare the performance of DC recovery methods by employing human experts, which will offer more convincing results than the objective IQA metrics used in this paper.

\bibliographystyle{IEEEbib}
\bibliography{ref}

\begin{thebibliography}{10}

\bibitem{USO:AC2DC:IEEETIP2006}
T.~Uehara, R.~Safavi-Naini, and P.~Ogunbona,
\newblock ``Recovering {DC} coefficients in block-based {DCT},''
\newblock {\em IEEE Transactions on Image Processing}, vol. 15, no. 11, pp.
  3592--3596, 2006.

\bibitem{ANR:DCT:IEEETComp1974}
N.~Ahmed, T.~Natarajan, and K.~R. Rao,
\newblock ``Discrete cosine transform,''
\newblock {\em IEEE Transactions on Computers}, vol. 23, no. 1, pp. 90--93,
  1974.

\bibitem{Shi-Sun:IVC2nd:Book2008}
Y.~Q. Shi and H.~Sun,
\newblock {\em Image and Video Compression for Multimedia Engineering},
\newblock CRC Press, 2nd edition, 2008.

\bibitem{JPEG}
ISO/IEC,
\newblock ``Information technology -- {Digital} compression and coding of
  continuous-tone still images: Requirements and guidelines,'' ISO/IEC 10918-1
  (JPEG), 1994.

\bibitem{Weng-Preneel:DC-Encryption:SIGMAP2007}
L.~Weng and B.~Preneel,
\newblock ``On encryption and authentication of the {DC DCT} coefficient,''
\newblock in {\em SIGMAP 2007 -- Proceedings of 2nd International Conference on
  Signal Processing and Multimedia Applications}, 2007, pp. 375--379.

\bibitem{Li:PVEA:IEEETCASVT2007}
S.~Li, G.~Chen, A.~Cheung, B.~Bhargava, and K.-T. Lo,
\newblock ``On the design of perceptual {MPEG}-video encryption algorithms,''
\newblock {\em IEEE Transactions on Circuits and Systems for Video Technology},
  vol. 17, no. 2, pp. 214--223, 2007.

\bibitem{Tang:MPEGEncryption:ACMMM96}
L.~Tang,
\newblock ``Methods for encrypting and decrypting {MPEG} video data
  efficiently,''
\newblock in {\em Proceedings of 4th ACM International Conference on
  Multimedia}, 1996, pp. 219--229.

\bibitem{Li:SPIC2008}
S.~Li, C.~Li, G.~Chen, N.~G. Bourbakis, and K.-T. Lo,
\newblock ``A general quantitative cryptanalysis of permutation-only multimedia
  ciphers against plaintext attacks,''
\newblock {\em Signal Processing: Image Communication}, vol. 23, no. 3, pp.
  212--223, 2008.

\bibitem{MeTriXMuX}
M.~Gaubatz,
\newblock ``{MeTriX MuX} visual quality assessment package,''
  http://foulard.ece.cornell.edu/gaubatz/metrix\_mux.

\bibitem{Wang:SSIM:IEEETIP2004}
Z.~Wang, A.~C. Bovik, H.~R. Sheikh, and E.~P. Simoncelli,
\newblock ``Image quality assessment: From error visibility to structural
  similarity,''
\newblock {\em IEEE Transactions on Image Processing}, vol. 13, no. 4, pp.
  600--612, 2004.

\bibitem{Wang:MS-SSIM:Asilomar2003}
Z.~Wang, E.~P. Simoncelli, and A.~C. Bovik,
\newblock ``Multi-scale structural similarity for image quality assessment,''
\newblock in {\em Conference Record of 37th Asilomar Conference on Signals,
  Systems and Computers}. 2003, vol.~2, pp. 1398--1402, IEEE.

\bibitem{Sheikh:IFC:IEEETIP2005}
H.~R. Sheikh, A.~C. Bovik, and G.~de~Veciana,
\newblock ``An information fidelity criterion for image quality assessment
  using natural scene statistics,''
\newblock {\em IEEE Transactions on Image Processing}, vol. 14, no. 12, pp.
  2117--2128, 2005.

\bibitem{Sheikh:VIF:IEEETIP2006}
H.~R. Sheikh and A.~C. Bovik,
\newblock ``Image information and visual quality,''
\newblock {\em IEEE Transactions on Image Processing}, vol. 15, no. 2, pp.
  430--444, 2006.

\end{thebibliography}

\end{document}